\documentclass[a4paper,11pt]{article}
\usepackage{pos}

\usepackage{bm}
\usepackage{cancel}
\usepackage{braket}
\usepackage{url}

\title{Internal structure of $T_{cc}$ and $X(3872)$ by using compositeness}

\author*{Tomona Kinugawa}
\author{Tetsuo Hyodo}

\affiliation{Department of Physics Tokyo Metropolitan University,\\
  Hachioji 192-0397, Japan}

\emailAdd{kinugawa-tomona@ed.tmu.ac.jp}
\emailAdd{hyodo@tmu.ac.jp}

\abstract{The internal structure of the near-threshold exotic hadrons, $T_{cc}$ and $X(3872)$, are studied by respecting the decay and coupled-channel contributions. The effective field theory model is introduced to calculate the compositeness, the probability of finding the hadronic molecular component. Applying the new interpretation scheme for the complex compositeness of unstable states and using the bare energies of the compact component estimated by the constituent quark model, we find that approximately 50 \% of the $T_{cc}$ wavefunction is occupied by the $D^{0}D^{*+}$ molecular component. This indicates that the structure of $T_{cc}$ is primarily dominated by the $D^0D^{*+}$ component, having the nearest threshold. However, other components, such as the isospin partner $D^{*0}D^{+}$ channels, also provide non-negligible contributions. On the other hand, about 90 \% of $X(3872)$ is composed of the $D^{0}\bar{D}^{0*}$ molecular component. This is because the non-composite bare state and the charged $D\bar{D}^{*}$ threshold are located relatively far from the $X(3872)$ and contribute very little to its internal structure.
}

\FullConference{The XVIth Quark Confinement and the Hadron Spectrum Conference (QCHSC24)\\
 19-24 August, 2024\\
 Cairns Convention Centre, Cairns, Queensland, Australia\\}

\begin{document}
\maketitle

\section{Introduction}

The numerous observations of exotic hadrons in recent years have stimulated active research on their internal structure~\cite{Hosaka:2016pey,Brambilla:2019esw}. For example, the Belle Collaboration reports the discovery of $X(3872)$ in 2003~\cite{Belle:2003nnu}, which is the first observed state among the $XYZ$ mesons. $X(3872)$ exists slightly below the $D^{0}\bar{D}^{*0}$ threshold and decays into $J/\psi \pi^{+}\pi^{-}$. Since that experiment, many candidates for exotic hadrons have been observed in the heavy sector. In particular, $T_{cc}$ is observed very near the $D^{0}D^{*+}$ threshold by LHCb Collaboration in 2021~\cite{LHCb:2021vvq,LHCb:2021auc}. Because $T_{cc}$ decays into $D^{0}D^{0}\pi^{+}$, it attracts much attention as a genuine exotic state with $cc\bar{u}\bar{d}$. 

Exotic hadrons can have not only a multiquark structure where four or more quarks are tightly bound by the strong interaction but also a hadronic molecular structure, the weakly bound state of hadrons~\cite{Guo:2017jvc}. Practically, the wavefunction of exotic hadrons is written as a superposition of possible structures, such as multiquark and hadronic molecular components. Here, we focus on the fraction of the hadronic molecular component. In this case, it is useful to introduce the compositeness $X$, the fraction of the hadronic molecular component in the wavefunction~\cite{Weinberg:1965zz,Kinugawa:2022fzn
}. 
Schematically, the compositeness represents the square of the coefficient of the molecular component $\ket{{\rm molecular}}$ in the bound state $\ket{B}$:
\begin{align}
\ket{B} &= \sqrt{X} \ket{{\rm molecular}} + \sqrt{Z}\ket{{\rm non\ molecular}}.
\label{eq:X-sch}
\end{align}
Here, we denote the non-molecular components, such as multiquark components, by $\ket{{\rm non\ molecular}}$. The fraction $Z = 1 - X$ is called the elementarity. Equation~\eqref{eq:X-sch} indicates that the compositeness $X$ can be regarded as the probability of finding the molecular component $\ket{{\rm molecular}}$ in the bound state $\ket{B}$:
\begin{align}
X &= |\braket{B|{\rm molecular}}|^{2}.
\end{align}
The compositeness allows us to quantitatively characterize the internal structure of exotic hadrons from the viewpoint of the hadronic molecule picture. For example, if $X > 50 \%$, the state is found to have a hadronic molecular-like structure, whose wavefunction is dominated by the hadronic molecular component. From this probabilistic nature, the compositeness has been adopted to quantitatively analyze the internal structure of exotic hadrons~\cite{Hyodo:2013nka,Oller:2017alp,vanKolck:2022lqz,Kinugawa:2024crb}. 

As in the $T_{cc}$ and $X(3872)$ cases, exotic hadrons are frequently observed near the threshold energy region. The phenomena in the near-threshold region are dictated by the low-energy universality~\cite{Braaten:2004rn,Naidon:2016dpf}. As a consequence of the low-energy universality, it is shown that the near-threshold bound states in the single-scattering system are usually composite dominant~\cite{Hyodo:2014bda,Hanhart:2022qxq,Kinugawa:2023fbf}. From this viewpoint, the near-threshold exotic hadrons are expected to have a molecular-like structure. However, the small binding energy is not the only property of exotic hadrons: they couple to the decay channel in the lower energy region and other scattering channels in the higher energy region. Due to the decay channel, exotic hadrons are unstable states with a finite decay width, which is qualitatively different from stable bound states. Furthermore, the higher coupled channel can affect the internal structure of exotic hadrons if the threshold energy of that channel is not far from the state. For example, $T_{cc}$ couples not only to the nearest threshold $D^{0}D^{*+}$ but also to the $D^{*0}D^{+}$ threshold. These two channels are mutually isospin partners, having tiny threshold energy differences. Therefore, the $D^{*0}D^{+}$ channel should be explicitly introduced to discuss the structure of $T_{cc}$. In this way, we need to consider the decay and coupled channel contributions to consider the internal structure of exotic hadrons. 

In this study, we investigate the internal structure of the representative exotic hadrons, $T_{cc}$ and $X(3872)$, using the compositeness. To calculate the compositeness, we introduce the non-relativistic effective field theory model. We discuss the internal structure of $T_{cc}$ and $X(3872)$ by focusing on not only the near-threshold nature of these states but also on the decay and coupled channel contributions.

\section{Compositeness of $T_{cc}$ and $X(3872)$}

Let us examine the internal structure of well-known near-threshold exotic hadrons, $T_{cc}$ and $X(3872)$. As introduced earlier, the states $T_{cc}$ and $X(3872)$ are observed as quasi-bound states located just below the $D^{0}D^{*+}$ and $D^{0}\bar{D}^{*0}$ thresholds, respectively. Their central values of the eigenenergies (binding energy $B$ and decay width $\Gamma$)
\begin{align}
E &= -B - \frac{\Gamma}{2}i,
\label{eq:eigenenergy}
\end{align}
have been measured in experiments, as reported in Ref.~\cite{LHCb:2021auc} and PDG~\cite{ParticleDataGroup:2024cfk}:
\begin{align}
T_{cc}&: E=-0.36-i0.024\ {\rm MeV}, 
\label{eq:Tcc-energy} \\
X(3872)&: E=-0.04 - i0.595\ {\rm MeV}. 
\label{eq:X3872-energy}
\end{align}
These binding energies are significantly smaller than the characteristic scale of the binding energy of hadrons of $\mathcal{O}(10)$ MeV. This makes $T_{cc}$ and $X(3872)$ representative examples of near-threshold quasi-bound states. The decay widths are observed as $0.048$ MeV for $T_{cc}$ and $1.19$ MeV for $X(3872)$. These values are smaller than the typical hadronic decay width of $\mathcal{O}(10)$ - $\mathcal{O}(100)$ MeV, and therefore the decay contribution is not always considered to discuss the internal structure of $T_{cc}$ and $X(3872)$ in the previous works. However, in Ref.~\cite{Kinugawa:2023fbf}, we find that the decay contribution to the compositeness is determined by the ratio of the decay width to the binding energy, not by the magnitude of the decay width itself. For shallow quasi-bound states, the ratio of the decay width to the binding energy can be large due to the small binding energy as summarized in Table~\ref{tab:Tcc-X}. In fact, the ratio $\Gamma/B$ is about 30 for $X(3872)$. Therefore, here, we explicitly take into account the decay width to consider the compositeness of $T_{cc}$ and $X(3872)$. In addition to the decay, systems have the scattering channels in the higher energy region than $T_{cc}$ and $X(3872)$. As mentioned in the introduction, the nearest scattering channels coupled to $T_{cc}$ and $X(3872)$ are $D^{0}D^{*+}$ and $D^{0}\bar{D}^{*0}$, respectively. Here, we call these channels the threshold channel. On top of these, there are $D^{0}D^{+}$ threshold for $T_{cc}$ and $D^{-}D^{+}$ threshold for $X(3872)$, which lie $1.41$ MeV and $8.23$ MeV above the threshold channels, respectively. These mass differences arise due to the isospin symmetry breaking, suggesting that the coupled channels may play a role in shaping the structure of $T_{cc}$ and $X(3872)$ as isospin partners of the threshold channels. Thus, both the decay width and channel coupling effects must be taken into account when investigating these exotic hadrons.

To calculate the compositeness of $T_{cc}$ and $X(3872)$, we adopt the framework of the non-relativistic effective field theory (EFT)~\cite{Braaten:2007nq,Kamiya:2016oao}. As discussed above, the decay and channel couplings can be important to consider $T_{cc}$ and $X(3872)$. Therefore, we construct a simple model by introducing the contributions of decay and channel couplings. The model describes the system where the two channel scatterings of $\psi_{1,2}$ and $\Psi_{1,2}$ couple to the discrete bare state $\phi$~\cite{Kinugawa:2023fbf}:
\begin{align}
\mathcal{H}_{\rm free}
&=
\frac{1}{2m_{1}}{\nabla}\psi_{1}^{\dag}\cdot {\nabla}\psi_{1} + \frac{1}{2m_{2}}{\nabla}\psi_{2}^{\dag}\cdot {\nabla}\psi_{2} 
+ \frac{1}{2M_{1}}{\nabla}\psi_{1}^{\dag}\cdot {\nabla}\Psi_{1} + \frac{1}{2M_{2}}{\nabla}\Psi_{2}^{\dag}\cdot {\nabla}\psi_{2} 
+ \frac{1}{2M}{\nabla}\phi^{\dag}\cdot {\nabla}\phi \nonumber \\
& \quad + \omega_{1}\Psi_{1}^{\dag}\Psi_{1}
+ \omega_{2}\Psi_{2}^{\dag}\Psi_{2}
+ \nu_{0}\phi^{\dag}\phi,
\label{eq:H-free} \\
\mathcal{H}_{\rm int}
&= g_{0}(\phi^{\dagger}\psi_{1}\psi_{2} + \psi^{\dagger}_{1}\psi^{\dagger}_{2}\phi)
+ g_{0}(\phi^{\dagger}\Psi_{1}\Psi_{2} + \Psi^{\dagger}_{1}\Psi^{\dagger}_{2}\phi).
\label{eq:H-int}
\end{align}
where $m_{1,2}$, $M_{1,2}$, and $M$ stands for the masses of $\psi_{1,2}$, $\Psi_{1,2}$, and $\phi$, respectively. $\omega_{1,2}$ are the energy of $\Psi_{1,2}$ with respect to $\psi_{1,2}$. We denote the threshold energy difference as $\Delta \omega = \omega_{1} + \omega_{2} > 0$. $\nu_{0}$ corresponds to the energy of the bare state, which is a real quantity. $g_{0}$ is the coupling constant among the bare state and the scattering states. In this model, we take $g_{0}$ to be complex in order to effectively introduce the decay channel contribution. With the complex coupling constant, the Hamiltonian becomes non-Hermitian whose eigenenergy $E$ can also be complex as in Eq.~\eqref{eq:eigenenergy}. This aligns with the fact that the eigenenergy of unstable states is complex. In this way, we can effectively take into account the decay channel contribution without explicitly calculating the three-body decay of $T_{cc}$ and $X(3872)$. For more detail on the model, please see Ref.~\cite{Kinugawa:2023fbf}.

In the model, we introduce the cutoff $\Lambda$ to avoid the divergence of the momentum integral in the loop function. The value of $\Lambda$ is chosen to be $140$ MeV by assuming the long-range interaction between $D$ mesons occurs through the pion exchange. In addition to the cutoff, we have two parameters $\nu_{0}$ and $g_{0}$ in the model. The bare state energy $\nu_{0}$ cannot be determined within the EFT framework, and a microscopic description is needed to estimate $\nu_{0}$. In this study, we adopt the energy of the compact $c\bar{c}$ and $cc\bar{u}\bar{d}$ states calculated by constituent quark models in Ref.~\cite{Karliner:2017qjm} and Ref.~\cite{Godfrey:1985xj}, respectively:
\begin{itemize}
\item $\nu_{0} = 7$ MeV for $T_{cc}$, and 
\item $\nu_{0} = 78.36$ MeV for $X(3872)$.
\end{itemize}
Once $\Lambda$ and $\nu_{0}$ are fixed, the coupling constant $g_{0}$ is uniquely determined to reproduce the binding energy $B$ and decay width $\Gamma$. We summarize the properties and model parameters of $T_{cc}$ and $X(3872)$ in Table~\ref{tab:Tcc-X}.

\begin{table}[tp]
\caption{The binding energy $B$, decay width $\Gamma$, threshold energy difference $\Delta\omega$ of the $T_{cc}$ and $X(3872)$ cases. We also show the parameters of the model, the bare state energy $\nu_{0}$ and cutoff $\Lambda$.}
\begin{center}
\begin{tabular}{|c|ccc|cc|}
\hline 
State & $B$ (MeV) & $\Gamma$ (MeV) & $\Delta \omega$ (MeV) & $\nu_{0}$ (MeV) & $\Lambda$ (MeV)  \\ \hline
$T_{cc}$ & 0.36 & 0.048 & 1.41 & 7 & 140 \\ 
$X(3872)$ & 0.04 & 1.19 & 8.23 & 78.36 & 140 \\ \hline
\end{tabular}
\end{center}
\label{tab:Tcc-X}
\end{table}

In this setup, the compositeness of $T_{cc}$ and $X(3872)$ can be calculated with the given model parameters and eigenenergy~\cite{Kinugawa:2023fbf}. For a multi-channel case, the compositeness of the threshold channel $X_{1}$, that of the coupled channel $X_{2}$, and elementarity $Z$ are defined. For the $T_{cc}$ [$X(3872)$] case, $X_{1}$ corresponds to the compositeness of $D^{0}D^{*+}$ ($D^{0}\bar{D}^{0*}$) channel and $X_{2}$ to that of $D^{*0}D^{+}$ ($D^{*-}D^{+}$) channel. Using the values in Table~\ref{tab:Tcc-X}, we obtain the compositeness $X_{1}$, $X_{2}$ and elementarity $Z$ of $T_{cc}$:
\begin{align}
X_{1} &= X_{D^{0}D^{*+}} = 0.541 - 0.007i \quad (T_{cc}), \label{eq:X1-Tcc} \\
X_{2} &= X_{D^{*0}D^{+}} = 0.167 + 0.003i, \quad (T_{cc})\label{eq:X2-Tcc} \\
Z &= 0.292 + 0.005 i \quad (T_{cc}). \label{eq:Z-Tcc}
\end{align}
Also, those of $X(3872)$ are calculated as
\begin{align}
X_{1} &= X_{D^{0}\bar{D}^{0*}} = 0.919 - 0.079i \quad [X(3872)], \label{eq:X1-X} \\
X_{2} &= X_{D^{*-}D^{+}} = 0.046 + 0.050 i \quad [X(3872)], \label{eq:X2-X} \\
Z &= 0.035 + 0.030 i \quad [X(3872)]. \label{eq:Z-X}
\end{align}
As seen in these results, the compositeness of unstable resonances is in general, complex. In the $T_{cc}$ case, the imaginary parts of $X_{1,2}$ and $Z$ are smaller than their real parts. In Ref.~\cite{Kinugawa:2023fbf}, it is shown that the effect of the decay becomes prominent if the ratio of the decay width $\Gamma$ to the quasi-binding energy $B$ increases. The small imaginary parts of $X_{1,2}$ and $Z$ of $T_{cc}$ can therefore be attributed to the small ratio of $\Gamma /B\sim 0.13$. From this result, we find that the decay contribution does not affect the compositeness of $T_{cc}$ very much. On the other hand, the imaginary parts of $X_{2}$ and $Z$ of $X(3872)$ are as large as the corresponding real parts. This is presumably related to the large ratio of $\Gamma /B\sim 30$ of $X(3872)$. This indicates the non-negligible decay contribution for the compositeness of $X(3872)$. 

The complex compositeness cannot be straightforwardly regarded as the probability. Therefore, with the complex compositeness shown above, we cannot directly discuss the internal structure of $T_{cc}$ and $X(3872)$. The interpretation of the complex compositeness has been discussed with various proposals of the interpretation schemes in Refs.~\cite{Hyodo:2011qc,Aceti:2012dd,Hyodo:2013nka,Aceti:2014ala,Guo:2015daa,Kamiya:2015aea,Kamiya:2016oao,Sekihara:2015gvw,Kinugawa:2024crb}. As a probabilistic interpretation scheme of the complex compositeness, here we adopt the prescription developed in Ref.~\cite{Kinugawa:2024kwb}. In this scheme, the internal structure of resonances is characterized by the following three quantities $\mathcal{X,Y,Z}$ defined by the complex compositeness $X$ and elementarity $Z$:
\begin{align}
\mathcal{X}&=\frac{(\alpha-1)|X|-\alpha|Z|+\alpha}{2\alpha-1}, \label{eq:calX} \\
\mathcal{Y}&=\frac{|X|+|Z|-1}{2\alpha-1}, \label{eq:calY} \\
\mathcal{Z}&=\frac{(\alpha-1)|Z|-\alpha|X|+\alpha}{2\alpha-1}. \label{eq:calZ} 
\end{align}
Here, $\mathcal{X}$, $\mathcal{Y}$, and $\mathcal{Z}$ are regarded as 
\begin{itemize}
\item[$\mathcal{X}$]: the probability to certainly finding the composite component;
\item[$\mathcal{Y}$]: the probability of uncertain identification; and 
\item[$\mathcal{Z}$]: the probability to certainly finding the elementary component.
\end{itemize}
$\mathcal{X}$ and $\mathcal{Z}$ correspond to the compositeness and elementarity for stable bound states, respectively. The newly introduced probability $\mathcal{Y}$ represents the uncertain property of resonances, inspired by the discussion in Ref.~\cite{Berggren:1970wto}. $\alpha$ is an arbitrarily real parameter, which determines the weight of the uncertain identification $\mathcal{Y}$ in the interpretation scheme. Here we use $\alpha \approx 1.1318$ which is determined by regarding the broad resonances as not interpretable by the compositeness. The details of the interpretation scheme can be found in Ref.~\cite{Kinugawa:2024kwb}. 

To use the interpretation scheme with Eqs.~\eqref{eq:calX}, \eqref{eq:calY}, and \eqref{eq:calZ} for the coupled-channel systems with $X_{1}, X_{2}$, and $Z$, we focus on the compositeness of the threshold channel $X_{1}$. Namely, we regard the compositeness $X=X_{1}$ as the fraction of the threshold channel component, and the other components are included in $Z=X_{2}+Z$, corresponding to the coupled channel compositeness and the elementarity in the coupled-channel systems. From the complex compositeness from Eq.~\eqref{eq:X1-Tcc} to Eq.~\eqref{eq:Z-X}, the probabilities $\mathcal{X,Y,Z}$ of $T_{cc}$ and $X(3872)$ are obtained as
\begin{align}
\mathcal{X} &= 0.537, \quad \mathcal{Y} = 0.008, \quad \mathcal{Z} = 0.456, \quad (T_{cc}),\\
\mathcal{X} &= 0.890, \quad \mathcal{Y} = 0.028, \quad \mathcal{Z} = 0.081, \quad [X(3872)].
\end{align}
In both cases, the compositeness $\mathcal{X}$ is the largest. This result reflects the shallow quasi-bound nature of $T_{cc}$ and $X(3872)$. Furthermore, $\mathcal{Y}$ is smaller than $\mathcal{X}$ and $\mathcal{Z}$, which indicates that the internal structure of these states can be clearly identified with small uncertainty. This is because the imaginary parts of the complex $X_{1}, X_{2}$, and $Z$ are as small as $10^{-2}$. By focusing on $T_{cc}$, the compositeness $\mathcal{X}$ is the largest, but with the non-negligible $\mathcal{Z}$. This is explained by the sizable magnitude of $X_{2}$ and $Z$ in Eqs.~\eqref{eq:X2-Tcc} and~\eqref{eq:Z-Tcc}, which are now included in the non-threshold channel components $\mathcal{Z}$. Because the threshold energy difference $\Delta\omega$ of the $T_{cc}$ system is relatively smaller (see Table~\ref{tab:Tcc-X}), the coupled channel compositeness $X_{2}$ is obtained as a comparable magnitude with $X_{1}$. The sizable magnitude of $Z$ is induced by the contribution of the purely non-composite bare state having $\nu_{0}=7$ MeV comparable with the typical energy scale $\Lambda^{2}/(2\mu) \sim 10$ MeV. From this result, we find that $T_{cc}$ is threshold channel dominant but with non-negligible contributions from the coupled channel and non-composite components. On the other hand, the compositeness $\mathcal{X}$ of $X(3872)$ is ten times larger than $\mathcal{Y}$ and $\mathcal{Z}$. Relatively small $Z$ in Eq.~\eqref{eq:Z-X} is a consequence of a large $\nu_{0}\sim 80$~MeV compared with the typical energy scale $\Lambda^{2}/(2\mu) \sim 10$ MeV. Also, the large threshold energy difference compared with the binding energy ($\Delta\omega/B\sim 200$) gives a similar magnitude of $X_{2}$. In this case, the non-composite and coupled-channel components have less effect on the internal structure of $X(3872)$. Therefore, $X(3872)$ can be regarded as an almost pure $D^{0}\bar{D}^{0*}$ molecule state. 

\section{Summary}

In this study, the internal structure of $T_{cc}$ and $X(3872)$ is studied using the compositeness. Based on the non-relativistic EFT model and the quark model estimation of the compact quark state energies, we calculate the compositeness of $T_{cc}$ and $X(3872)$ by taking into account the decay and coupled-channel contributions. We find that $T_{cc}$ is the $D^{0}D^{*+}$ molecular dominant state, but other components are also sizable. In contrast, $X(3872)$ is almost exclusively dominated by the $D^{0}\bar{D}^{0*}$ molecule state. This difference is considered to originate from both the threshold energy of the coupled channel and the bare state energy, compared with respect to the binding energy. 

\section*{Acknowledgments}

This work has been supported in part by the Grants-in-Aid for Scientific Research from JSPS (Grants
No.~JP23KJ1796, 
No.~JP23H05439, 
No.~JP22K03637, and 
No.~JP18H05402), 
and by JST, the establishment of university fellowships towards the creation of science technology innovation, Grant No. JPMJFS2139. 

\bibliographystyle{h-physrev3}

\end{document}